\documentclass[aps,pra,twocolumn,superscriptaddress,nofootinbib]{revtex4-1}
\usepackage{amssymb,graphicx,color,hyperref}
\usepackage[intlimits]{amsmath}
\usepackage[english]{babel}

\def\beq{\begin{equation}}
\def\eeq{\end{equation}}
\def\bea{\begin{eqnarray}}
\def\eea{\end{eqnarray}}
\def\barr{\begin{array}}
\def\earr{\end{array}}

\begin{document}

\title{Benefits of non-adiabatic quantum control in quantum computation through spin qubit systems}

\author{Nirupam Dutta}
\email[E-mail:]{nirupamdu@gmail.com}
\affiliation{Independent Researcher, 25B Shanti Ghosh Street, Kolkata, West Bengal, India}

\date{\today}

\begin{abstract}
This is evident that the controllable quantum systems can be the reliable building blocks for  Quantum computation. In reality we are witnessing the progress towards making the idea tractable enough, though optimistic but the threshold is not very near to us. The dawn of quantum computation has begun. In the future, we hope to see a full fledged operationally stable quantum computer which can solve the problems beyond the scope of classical digital computers. We may call it quantum supremacy. Nevertheless, we should not forget that there are problems which demand classical computers to be in the game for a better performance in comparison to the same through quantum devices. In the current stage of computing technology, the most beneficial area is nothing but an hybrid approach and that is  for no doubt will reign the market for the next five to ten years. This hybrid aspect has several directions such as simulating quantum computation on a classical computer. Keeping both the aspect, computation through real physical devices and simulation on a classical computer by accessing available quantum computers for cloud computing, some advantages have been discussed in this article which will be elaborated as well in future articles. These advantages are inherent if we can achieve proper non-adiabatic control over the spin system in the laboratory. Otherwise these aspects can always be simulated by using quantum algorithms to see whether they can be useful in comparison to a purely classical computing machine. This is no doubt a new window for progress in the direction of quantum computation. 
\end{abstract}

%\pacs{}

\maketitle
We need controllable quantum systems in order to employ their quantum mechanical properties to perform computational tasks \cite{zero, one, two} that are either beyond the scope of any classical computer or quicker in processing information in comparison to the same performed on a classical computing machine \cite{three, four}. This agenda is ruling the current state of the art in the domain of computation. A large number of theorists and experimentalists at various organisations are participating in this race for achieving the so called quantum supremacy  \cite{five, six, seven, eight}. Physical systems at the atomic and subatomic scale are quite difficult to control in order to manipulate them according to our computational needs.  For example a two level quantum system which is  the most basic text book topic of quantum mechanics can not be prepared easily in the laboratory. It has already been four decades that scientists are advocating in favour of quantum computation as in many ways it can revolutionise the subject of computation and perhaps can push the current technological horizon. Theoretical understanding has already made us able to realise that the quantum superposition, interference and entanglement are three important pillars \cite{nine} for making this dream a reality. In today's scenario, many have achieved creating qubits in various different systems under sophisticated laboratory conditions.
\\

Quantum computation, we all know is different from classical computation in the following way. Instead of classical beats which are either on or off like a switch in the classical circuit, here the basic component is a superposed state of a two level system which we name as qubit. But unless there is entanglement among the qubits, they are no where more efficient than classical bits as it somehow creates the technical essence of parallel computing. With the help of the qubits, like we do for a classical computing device, we can create gates for performing basic operations through qubits and at the end a measurement of this physical process though destroys the superposition but strategically designing the circuit for constructive interference we assure the desired results. This is the general process of gate based quantum computation but this is not the only mode of quantum computation. There are quantum algorithms written for collection of qubits which do not require anything like a gate structure for computing complex problems like optimisation problems or factorisation problems. Quantum annealing \cite{ten, nishi1, nishi2} is one of the examples which deals with adiabatic quantum computation to solve optimisation problems. Almost decade later the D-WAVE device \cite{king} is successfully handling various tasks and has already been updated to a controlled device dealing with more than thousand qubits.

There are few types of quantum mechanical systems which have been successfully employed so far to create qubits in the laboratory. These are for example superconducting qubits, trapped ions, laser controlled photons and controlled spin qubits. In order to achieve operationally functional computing devices which can run through different quantum algorithms, hundreds of thousands  of stable qubits are needed. In this context it is worth mentioning that this number is not actually close tho the number of qubits needed for achieving quantum supremacy of the computing device. On the other hand, these large numbers of qubits have to be entangled with each other in order to become clusters of logical qubits to be able to perform computation. It is evident that to perform different types of tasks based on different algorithms, the number of active quantum qubits are different but a sustainable correlation among them can not be assured trivially. The reason is simple and that is nothing but the measurement sensitivity of quantum systems.  

The most important thing that we still should keep in mind in this context is that these different types of protocols and schemes that I have discussed mostly employ the adiabatic control over the quantum systems. The basics of quantum annealing are strictly based on the adiabatic control over the system. The Hamiltonian of the system is strictly maintained within the slow rate of change such that it should not break the adiabatic condition. On the other hand, gate based quantum computation has also been developed through adiabatic control.

In this article I will talk about some benefits of nonadiabatic controls over quantum systems. In my demonstration I will first talk about physical spin systems in which formation of physical qubits can be tailored by non adiabatic evolution and in the same context, I will additionally discuss how nonadiabatic effects can give rise to the creation of several qubits almost in no time from parent qubits. These are important in real physical systems when they can be controlled within a confined region of space but this challenge obviously can be tackled by the experimentalist as far as the viability of the process is concerned. 

Furthermore, a full fledged computer is still a matter of waiting for the sufficient number of controllable qubits. Considering the state of the art, the recipe we present here for non adiabatic control  can be useful in simulating various schemes of quantum computation through a hybrid approach that combines the cloud computing facilities of available quantum computers and simulation of various methods of solving complex problems through specific quantum algorithms.

\vskip 2cm

\section{A. Qubit Multiplier}
\begin{figure}
\begin{minipage}{0.5\textwidth}
\includegraphics[width=0.9\textwidth]{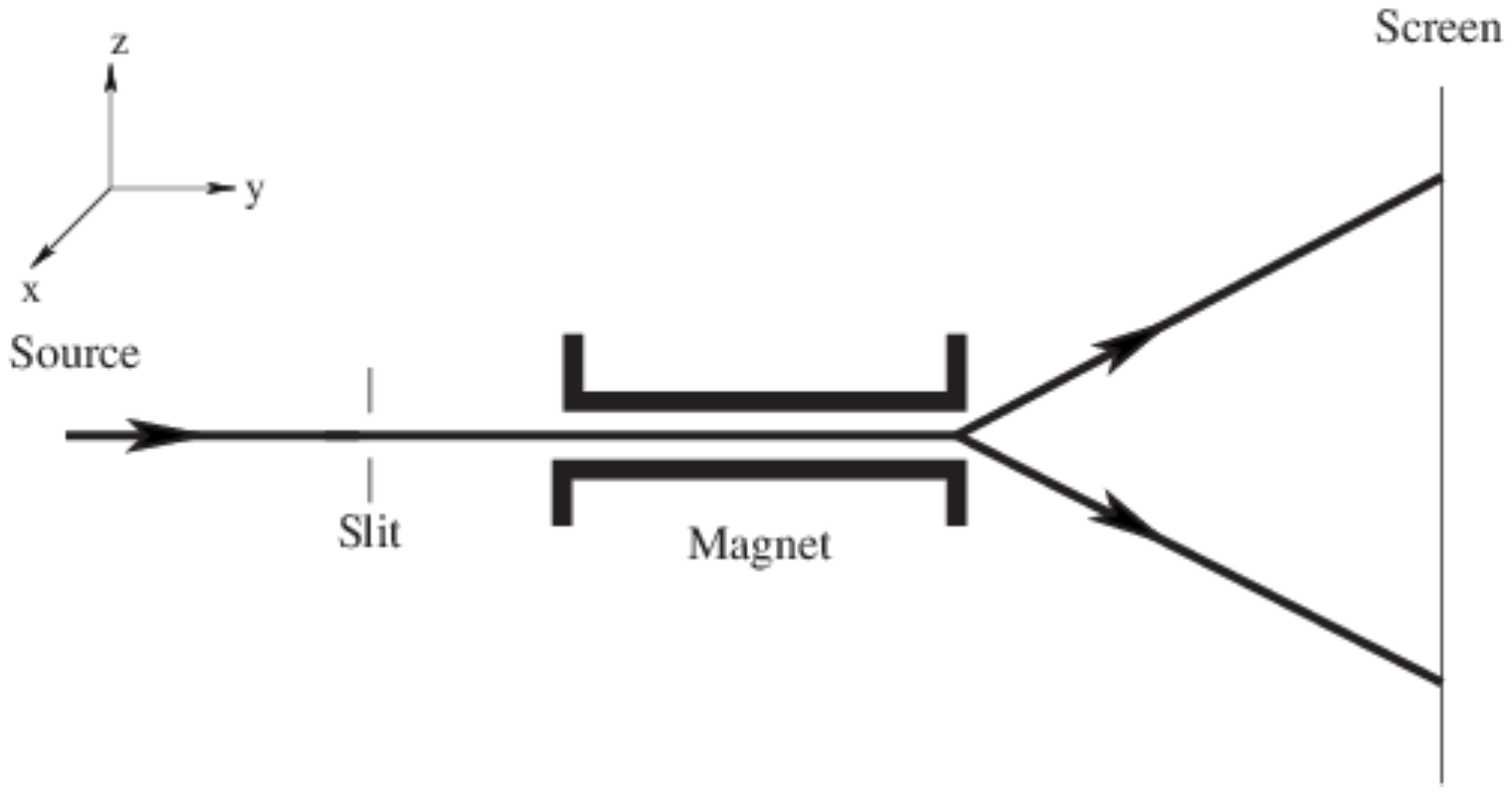}
\caption{\small This is a schematic diagram of a Stern-Gerlach device which shows that the inhomogeneous magnetic field actually separates two spin states in the up and down direction.}
\label{fig:one}
\end{minipage}
\end{figure}

Let me start with a two level system for example a physical spin qubit. A spin qubit can always be prepared through any available mechanism. Now if we allow the qubit $a_{0}|0\rangle + a_{1} |1\rangle$ to pass through an inhomogeneous time dependent or independent magnetic field, the evolved state should be  determined through the nature of the evolution. We are using $0$, $1$ to denote the up and down spin along the $z$ axis of the specified coordinate system \ref{fig:one}. If the system evolves through an adiabatic process the final state remains one qubit but the state will be entangled with the wave functions in the up and down directions. This phenomena is known as path spin entanglement \cite{home}. The final state $|\Psi(t)\rangle$ can be expressed as,

\begin{equation}
|\Psi(t)\rangle = a_{0}e^{-i\omega_{0}t}|0\rangle \otimes \phi_u+ a_{1}e^{i\omega_{0}t} |1\rangle  \otimes \phi_d\label{adia1}
\end{equation}

By path spin entanglement, I mean the spatial wave functions $ \phi_u$  and $ \phi_d$ corresponding to upward and downward direction along the $z$ axis are coupled to the up and down spin states. This is evident from the above equation that spatial and spin part of the state are not product separable. For an operationally and formally ideal situation of the Stern-Gerlach set up \ref{fig:one}, the up and down part of the wave function can be well separated.
In that case, if an ensemble of $N$ number of spin half particles undergo this magnetic field configuration, the up and down spins will be accumulated within a small region of the space in the upward and downward directions respectively. The probability of having the up spin in the spatially upward direction is given by,

\begin{equation}
P_{0,u} = {a_0}^2 \int { \phi_{u}}^{2} dX .
\end{equation}
Where $X$ represents spatial variables.

So the total number of accumulated particles $N_{0,u}$ in the state $ |0 \rangle $ in the upward direction is $N$ times the probability $P_{0,u}$. Similarly total number of accumulated particles $N_{0,d}$ having the down spin $|1\rangle $ in the downward direction is $NP_{1,d}$ as  the probability of having down spin in the downward direction is given by,

\begin{equation}
P_{1,d} = {|{a_1}|}^2 \int { \phi_{d}}^{2} dX
\end{equation}

Actually each of the $N_{0,u}$ number of particles are in the up state  and at the same time the cluster itself can be used as one logical bit $|0\rangle$. The same argument applies to the cluster of particles which are confined in the downward direction.

\subsection{A. 1. Converter Oracle}
This adiabatic evolution and path spin entanglement has an interesting feature in the context of computations through the qubits. Suppose we start with a qubit state and with one step boosting through regular ideal S-G set up, we can create well separated two streams of two different states. Now,  among the two streams, if we allow only the upward one to be released, when considered as a final output, the manipulation results in conversion of a superposed state to a pure state which is equivalent to a classical bit.
\begin{figure}
\begin{minipage}{0.5\textwidth}
\includegraphics[width=0.9\textwidth]{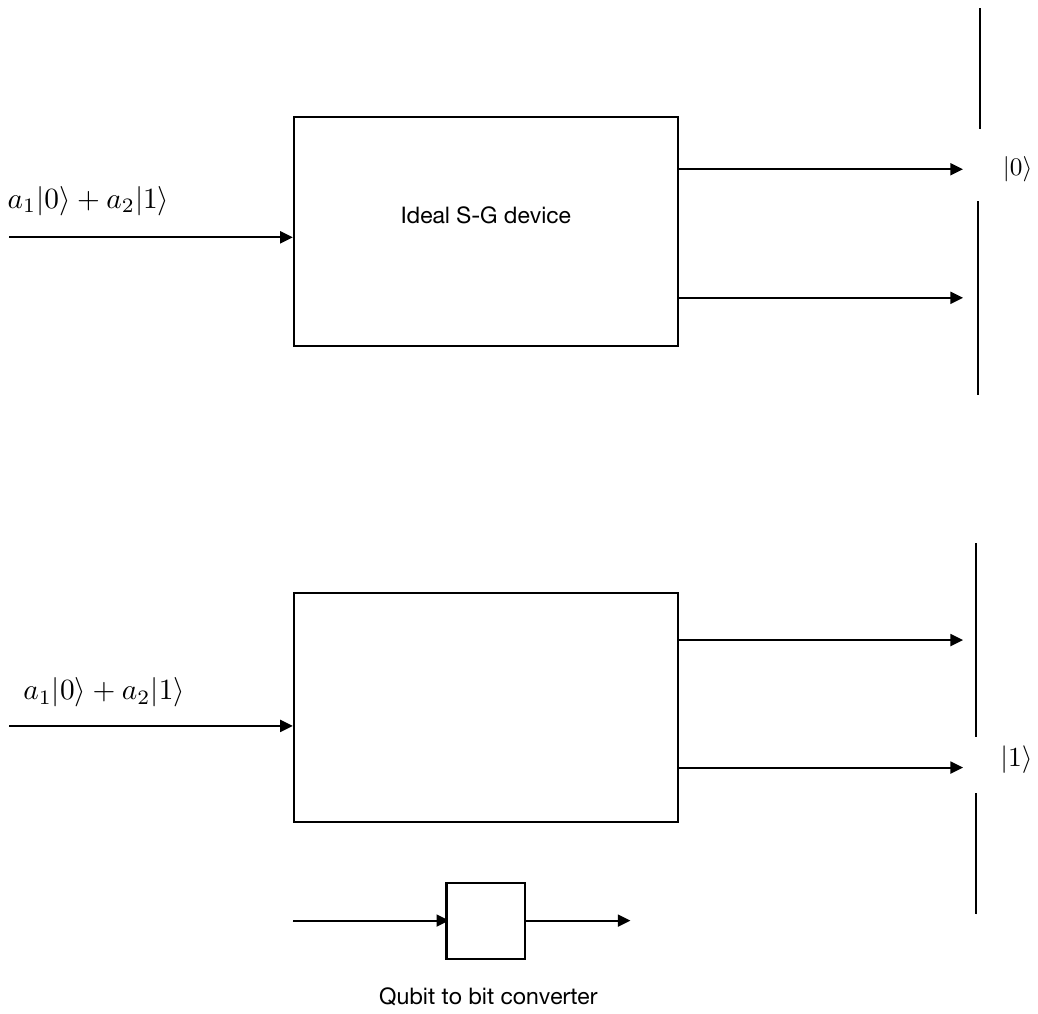}
\caption{\small This is a schematic diagram of the converter which can transform a superposed state toa pure state with the help of ideal S-G apparatus.}
\label{fig:two}
\end{minipage}
\end{figure}
 The entire process in this case serves like an oracle which changes an input qubit to an output which serves as a classical bit. A reverse process can always restore to the initial state and hence can keep the unitarity unviolated through the process. This oracle can be used as a gate that transforms  quantum to classical beats and can be useful in various algorithms as a component of an integrated circuit to solve specific computational problems.
\subsection{A. 2. The non-adiabatic multiplier }
A very different situation arises when the system evolves through a non-adiabatic evolution.  For this purpose, the Hamiltonian of the spin system needs to be  changing quite fast with respect to time. How fast it should be is determined with respect to the a characteristic time scale $\tau$ of the system which is the inverse of the energy gap between the Zeeman levels, I mean the energy states of the two level system. If the characteristic time $\tau$ is much less than the time needed for the unit amount change of the Hamiltonian, the system evolves adiabatically otherwise it evolves non-adiabatically. The precise criteria for non-adiabatic evolution is as follows \cite{aharonov, amin, tong}.

\begin{equation}
\Big|\frac{\langle 0(t)|\dot{H}(t)| 1(t)\rangle}{(E_{0}-E_{1})^2}\Big|<<1. \label{condn}
\end{equation}
$|0(t)\rangle$ and $|1(t)\rangle$ are the instantaneous eigenstates of $H(t)$ and $E_0$ and $E_1$ are the corresponding energy eigenvalues.\\

There can be many different possible magnetic field configurations, which can violates this condition and serve our purpose. For a simple demonstration, let me consider a very well known configuration which has been understood previously in the context of spin states in other areas of physics. Here we will be enjoying a few advantages for the purpose of quantum computing. The considered field configuration is a rotating magnetic field and in order to have the desired outcome, I will present a specific case when a rotating field component will be added to a formally and operationally ideal Stern-Gerlach setup. The rotating field is confined in the X-Y plane only and it rotates about the Z axis. Another method of violating adiabatic condition was presented in one of our previous articles in the context of evolution of spin states when there is azimuthal inhomogeneity inside the S-G set up \cite{dutta}. In both the cases the mathematical construction of the problem is same. This field strength has to be decided in such a way that the time scale of spin states dynamics is much smaller than that of the spatial wave function. Then, We can represent the Hamiltonian in the following way.
\begin{eqnarray}
H(t)\nonumber = \gamma B(\sigma_x \sin\theta \cos\phi (t) \\+ \sigma_y \sin\theta \sin \phi (t)+ \sigma_z \cos\theta) \nonumber \\=  \omega_0(\sigma_x \sin\theta \cos\omega t \\ \nonumber+ \sigma_y \sin\theta \sin \omega t + \sigma_z \cos\theta).
\end{eqnarray}
 Where $\gamma$ is the gyromagnetic ratio and $\sigma_x$,  $\sigma_y$,  $\sigma_z$ are three different Pauli spin matrices in the chosen orientation of the total spin $S$. The rotating components makes the field rotating in the x-y plane with a frequency $\omega$. $\theta$ is the angle between the z-component of the magnetic moment ($\mu =\gamma S $) and the magnetic field $\vec{B}$ and $\phi(t)$ is the azimuthal angle that magnetic field makes in the $X-Y$plane considered here. The matrix form of the Hamiltonian is given below.

\begin{eqnarray*}
\begin{aligned}
 H(t)= \frac{\omega_0}{2}\left(
\begin{array}{cc}
\cos\theta & e^{-i\omega t} \sin\theta  \\ 
e^{i\omega t} \sin\theta  & - \cos\theta \\ 
\end{array}
\right)
\end{aligned}
\end{eqnarray*}   
 The quantity $\omega_0$ is the Larmor frequency of the magnetic moment of the spin$1/2$ particle.

The instantaneous Eigenstates of the time dependent Hamiltonian can be expressed as,

\[
|0(t)\rangle =
\begin{pmatrix}
e^{-i\omega t/2}\sin\theta/2\\
e^{i\omega t/2}\cos\theta/2
\end{pmatrix}
\]
and
\[
|1(t)\rangle =
\begin{pmatrix}
e^{i\omega t/2}\sin\theta/2\\
-e^{-i\omega t/2}\cos\theta/2
\end{pmatrix}
\].

Plugging these two states in the equation \ref{condn}, we arrive to the condition for adiabatic evolution as,
\begin{equation}  
\frac{\omega}{2\omega_0}\sin\theta<<1.\label{adia}
\end{equation} 

This equation tells us that whether the evolution is adiabatic or not depends on three different parameters: $\omega_{0}$, $\omega$ and $\theta$. One can employ this freedom to break the adiabatic condition and in that case the evolution of spin states will be determined by solving the Schrödinger equation.

\begin{figure}
\begin{minipage}{0.5\textwidth}
\includegraphics[width=\textwidth]{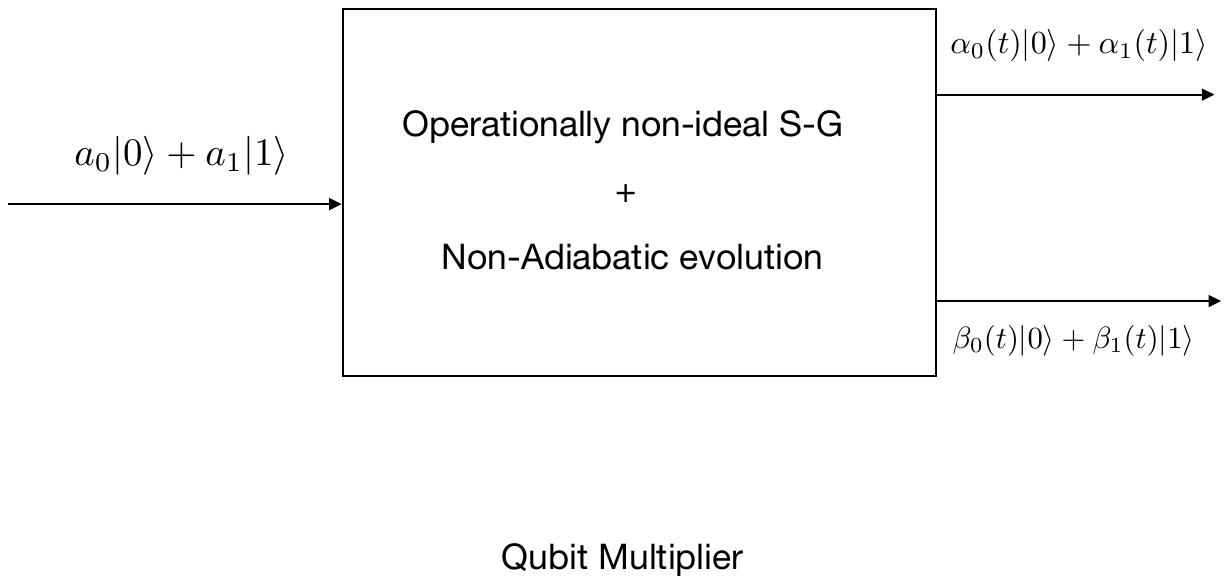}
\caption{\small This is a schematic diagram of the oracle "qubit multiplier" which is based on the non-adiabatic control inside a formally and operationally ideal Stern-Gerlach device.}
\label{fig:three}
\end{minipage}
\end{figure}

Now let me come back to the story that we have started in this section. Suppose the initial qubit is a superposition of two spin states $|0\rangle$ and $|1\rangle$. The moment they enter into the magnetic field, the up and down states will be coupled to the spatial up and down component of the wave function and will evolve according to the time dependent Hamiltonian. The interaction Hamiltonian $H$ does not contain any spatial part hence the time dependence will affect the spin part only. The spatial inhomogeneity of the basic ideal S-G setup will help separating the wave functions in the upward and downward direction along the Z axis. So we just need to study the evolution of up and down spin states under the influence of the additional rotating component.

We have to calculate the evolution of the up state which will be deflected in the upward direction and for the down state which will be deflected in the downward direction. For both purposes we have to solve the Schrödinger equation for the time dependent Hamiltonian with two different initial conditions with the help of the following instantaneous Eigenstates.
Let's write down the evolved state in the upward direction in the following way,

\begin{equation}
\Psi_{u,z} = |s_{zu}(t)\rangle \otimes \Phi_{u}.
\end{equation}
Where,
\begin{equation}
|s_{zu}(t)\rangle = \alpha_{0}(t) |0(t)\rangle+ \alpha_{1}(t)|1(t)\rangle \nonumber
\end{equation}
Now, We see that the $N_u$ number of particles forms the cluster of qubits which will be accumulated in the upward direction. On the other hand the $N_d$ number of particles will form a cluster of qubits in the downward direction. The state can be written as,

\begin{equation}
\Psi_{d,z} =  |s_{zd}(t)\rangle \otimes \Phi_{d}.
\end{equation}
Where,
\begin{equation}
|s_{zd}(t)\rangle = \beta_{0}(t) |0(t)\rangle+ \beta_{1}(t)|1(t)\rangle \nonumber
\end{equation}

Now plugging this in the Schrödinger equation with the initial conditions $a_{0}(t=0) = 1 $ for the upward stream and $a_{1}(t=0) = 1 $ for the down stream, we are left with the following solutions for $\alpha_{0}$, $\alpha_{1}$, $\beta_{0}$ and $\beta_{1}$ as follows.

\begin{eqnarray}
\begin{aligned}
\alpha_{0}(t)&= \cos \frac{\bar{\omega}t}{2}-i \frac{\omega_0-\omega \cos\theta}{\bar{\omega}} \sin\frac{\bar{\omega}t}{2}\\
\alpha_{1}(t)& = i \frac{\omega \sin\theta}{\bar{\omega}}\sin \frac{\bar{\omega}t}{2}\\
\beta_{0}(t)& = i \frac{\omega \sin\theta}{\bar{\omega}}\sin \frac{\bar{\omega}t}{2}\\
\beta_{1}(t)& = \cos \frac{\bar{\omega}t}{2}+i \frac{\omega_0-\omega \cos\theta}{\bar{\omega}} \sin\frac{\bar{\omega}t}{2}
\end{aligned}
\end{eqnarray}

Where,  $\bar{\omega} = \sqrt{\omega_0^2+\omega^2-2\omega_0 \omega \cos\theta}$\\

The first advantage of this nonadiabatic process is that we are able to create two different logical qubits through two different clusters of physical qubits. This is like an oracle which doubles the number of logical qubits  just by giving one order of non-adiabatic boosting through the time dependent magnetic field inside the ideal S-G setup. With such $n$ number of non adiabatic booster the logical qubits can be scaled by a factor of $2^n$.
\\

The second advantage is that we can create many basic one qubit logic gates through the non-adiabatic control. Let us discuss just a few among these gates and the associated values of the control parameters. There are four different parameters such as  $\omega_0$,   $\omega$,   $\theta$ and $t$ which can be adjusted to achieve the desired output. The scope of exploration is almost unlimited but let me just show one particular choice of these free parameters and how we can achieve three different one qubit quantum logic gates through this. 

Lets start with a particular choice which is $\cos\theta = \frac{\omega_0}{\omega}$. The ratio of the two frequencies have to be chosen in such a way so that it does not satisfy the adiabatic condition \ref{adia}. It is very much possible as there is a big domain for which the ratio $\frac{\omega_0}{\omega}$ can be fixed accordingly.\\

 \textbf{Case 1. Quantum Not Gate:}
Suppose the incoming state to the S-G setup is $a_{0}|0\rangle + a_{1} |1\rangle$ The evolved state in the upward direction becomes $a_{1}|0\rangle + a_{0} |1\rangle$ at some instant $t= \tau_{not}$ for which 
\begin{equation}
\cos \frac{\bar{\omega}\tau_{not}}{2} = a_1 \hskip 0.3cm \text{and} \hskip 0.3cm  i\sin \frac{\bar{\omega}\tau_{not}}{2} = a_0
\end{equation}
For this purpose the time dependent Hamiltonian should change up to $H(\tau_{not})$ from  $t = 0$ to $t = \tau_{not}$ and should remain fixed afterwards. The qubit at the final instant can be treated as the output of a quantum NOT gate and can be utilised accordingly. Under this circumstance, the qubits in the downward direction is the same as the input qubit. This operation is equivalent to an Identity operation on the input state \\

 \textbf{Case 2. The Z Gate:}
 With the same value of $\theta$, if we consider the final state in the upward direction at some instant $\tau_z$ so that it satisfies the following condition.
\begin{equation}
\cos \frac{\bar{\omega}\tau_{z}}{2} = a_0 \hskip 0.3cm \text{and} \hskip 0.3cm  i\sin \frac{\bar{\omega}\tau_{z}}{2} = -a_1
\end{equation}

 In this case the final outcome in the upward direction is $a_{0}|0\rangle - a_{1} |1\rangle$.
Obviously, the Hamiltonian in this case shall remain unchanged from $t=\tau_z$ with the final value $H(\tau_{z})$.\\

 \textbf{Case 3. The Hadamard Gate:}
 For the same value of $\theta$, if we impose the following condition,
 
 \begin{equation}
\cos \frac{\bar{\omega}\tau_{H}}{2} =\frac {a_0 + a_1}{\sqrt{2}}  \hskip 0.3cm \text{and} \hskip 0.3cm  i\sin \frac{\bar{\omega}\tau_{H}}{2} = \frac {a_0 - a_1}{\sqrt{2}}
\end{equation}
The final state in the upward direction is $a_0 \frac{|0\rangle + |1\rangle}{\sqrt{2}} + a_0 \frac{|0\rangle - |1\rangle}{\sqrt{2}}$. This is the output of a single qubit Hadamard gate.

\section{B. Non-adiabatic control over multi qu-bit entangled states}
So far, we have talked regarding the non-adiabatic control through the rotating time dependent magnetic field inside a formally and operationally ideal S-G apparatus. In this section we will discuss the control we can have over multi qubit entangled states which is one of the important ingredients for quantum computation. I won't be making this section prolonged as in an upcoming article I will show preparation of various many qubit quantum gates and specific controls for introducing algorithms to solve complex problems. Let me give an introductory description along with a few important hints that can create some ambition for future study in the domain of non-adiabatic quantum control. 

Instead of S-G set up, here we consider Mach-zehnder setup in conjunction with an S-G device for the purpose of creating entanglement among different multi-particle  qubits \cite{pan}. This is actually done through the mechanism of swapping of path-spin entanglement to spin spin entanglement. The issue has been well studied previously. I am not going into the details of the mechanism as we are trying to emphasise on the improvement that we are bringing in through the application of an additional time dependent rotating magnetic field. Through a specific process first the mechanism creates two particle (particle $1$ and $2$) states which are coupled to a pseudo path like spatial variable and later a third particle 3 is introduced which previously has not interacted with the two particles. The swapping then creates various entanglement between 2 and 3. Intuitively, the process is the following. Particles ‘2’ and ‘3’ are kept separated without allowing them to interact throughout the process. Now the scheme allows the particle ‘1’ to interact independently with the particles ‘2’ and ‘3’, without the need of having ‘2’ and ‘3’ in the near vicinity of each other.

Now the qubit undergoes through two different S-G apparatus and these devices provide an unitary transformation /cite. Upon measurement mentioned on the state of the particle $1$ we can create two different kinds of entangled states which are given below.\\

$1)$. $a |00\rangle_{23} + b  |11\rangle_{23}$

$2)$. $a |00\rangle_{23} -b  |11\rangle_{23}$
\\

But if we use the non adiabatic control over the outgoing state of the Mach- Zehender state through a time dependent rotating magnetic field, we can create four different types of entangled states in the two particle tensor product space of particle $2$ and $3$. Furthermore, by adjusting the parameter values, we can create various two qubits gates in this context. In this article we are not going into the details of those gates and quantum circuits that can be employed to create various quantum algorithms for solving complex issues. This issue I leave for a different article in near future.

\section{Discussion}
My purpose in this article is to introduce the key idea of non adiabatic control and the various benefits that it can offer for computing. Now implementation of the control in the form of a physical device needs necessary engineering and that might suffer from several difficulties in the laboratory. Nevertheless the procedure opens up new possibilities for integrated devices which can be useful in solving various complicated problems with the help of proper algorithms. On the other hand, a mathematical simulation of non-adiabatic control can be a good approach if it is implemented through quantum algorithms running on basic quantum computers that are available for cloud computing. In a different article we will show a few such algorithms and how they can be useful in solving various computations in the simulation of many body physics and chemistry as well as in the domain optimization problems.

\end{document}